\documentclass[fleqn,usenatbib]{mnras}

\usepackage{newtxtext,newtxmath}

\usepackage[T1]{fontenc}
\usepackage{ae,aecompl}

\usepackage{graphicx}	
\usepackage{amsmath}	

\newcommand{\Msun}{M$_{\sun}$}
\newcommand{\Rpl}{\,R_\mathrm{pl}}
\newcommand{\Mpl}{\,M_\mathrm{pl}}
\newcommand{\kms}{\,km~s$^{-1}$} 
\newcommand{\amuse}{\textsc{amuse}}
\newcommand{\phf}{\textsc{ph4}}
\newcommand{\bridge}{\textsc{bridge}}

\title[SC survival in the slingshot scenario.]{Stirred, not shaken: Star cluster survival in the
    slingshot scenario
}

\author[D. R. Matus Carrillo et al.]{
D. R. Matus Carrillo,$^{1}$\thanks{E-mail: dimatus@udec.cl (DMC)}
M. Fellhauer,$^{1}$
T. C. N. Boekholt, $^{2}$
A. Stutz,$^{1}$\newauthor
and M. C. B. Morales Inostroza$^{1}$
\\
$^{1}$ Departamento de Astronom\'{i}a, Universidad de Concepci\'{o}n,Casilla 160-C, Concepci\'{o}n, Chile\\
$^{2}$Rudolf Peierls Centre for Theoretical Physics, Clarendon Laboratory, Parks Road, Oxford, OX1 3PU, UK \\
}

\date{Accepted XXX. Received YYY; in original form ZZZ}

\pubyear{2023}

\begin{document}
\label{firstpage}
\pagerange{\pageref{firstpage}--\pageref{lastpage}}
\maketitle

\begin{abstract}
    We investigate the effects of an oscillating gas filament on the dynamics
    of its embedded stellar clusters.
    Motivated by recent observational constraints, we model the host  gas
    filament as a cylindrically symmetrical potential, and the star cluster as
    a Plummer sphere.
    In the model, the motion of the filament will produce star ejections from
    the cluster, leaving star cluster remnants that can be classified into four
    categories: a) Filament Associated clusters, which retain most of their
    particles (stars) inside the cluster and inside the filament; b) destroyed
    clusters, where almost no stars are left inside the filament, and there is
    no surviving bound cluster; c) ejected clusters, that leave almost no
    particles in the filament, since the cluster leaves the gas filament; and
    d) transition clusters, corresponding to those clusters that remain in the
    filament, but that lose a significant fraction of particles due to
    ejections induced by filament oscillation.
    Our numerical investigation predicts that the Orion Nebula Cluster is in
    the process of being ejected, after which it will most likely disperse into
    the field.  
    This scenario is consistent with observations which indicate that the Orion
    Nebula Cluster is expanding, and somewhat displaced from the Integral
    Shaped Filament ridgeline.

\end{abstract}




\section{Introduction}\label{sec:intro}

The majority of stars form in filamentary gas structures in molecular
clouds~\citep{andre_2010}.
Star clusters are no exception.
In the early phases, the star clusters are gas-dominated~\citep{lada_2003}, and
the gas is arranged in simple structured filaments that can be approximately
modelled as cylinders~\citep{stutz_2016}.
The key question here is how the gas properties may affect the embedded cluster
properties and dynamics.
The nearest \citep[$\sim$400 pc, ][]{kounkel_2018, stutz_2018a}, and arguably
best-studied embedded cluster is the Orion Nebula Cluster
\citep[ONC,][]{hillenbrand_1997,hillenbrand_1998}, which is forming within the
massive Integral shaped Filament \citep[ISF,][]{bally_1987}.
These star and gas structures are embedded in the massive ($M\sim10^5$\Msun)
Orion A molecular cloud \citep[OMC,][]{bally_1987}.

The OMC has a mass of $1.1\times10^5$~\Msun \citep{hartmann_2007}, distributed
along an extension of $\sim90$ pc \citep{grosschedl_2018}.
This structure is composed of many smaller filaments and clumps, with over 100
individual condensations identified \citep{bally_1987}.
Of these structures, one of the most striking is the ISF, named as such thanks
to its distinctive shape, which corresponds to the northern part of the OMC.

In the middle of the ISF, the ONC is forming, as mentioned above.
The ONC is the brightest and most prominent stellar structure in Orion A, and
it's nebulosity is visible with the naked eye.
The estimated ONC total mass is $\sim1000$~\Msun~\citep{da_rio_2012,stutz_2018}.
The mean stellar mass is $\sim$~0.7\Msun~\citep{hillenbrand_1997,
takemura_2021}, with individual stellar masses that range from below the
hydrogen burning limit to $\sim$~33~\Msun\citep{hillenbrand_1997, balega_2014}.
In isolation, a bound star cluster will evolve towards a spherical shape, due to
the gravitational interaction of the stars.
The ONC, being partially embedded inside the ISF, is not isolated, and its
stars are moving under the influence of the gas.
As a consequence, it is not circularly symmetrical, and it is elongated 
similar to the gas distribution in the region \citep{hillenbrand_1998},
with an ellipticity between 0.3 to 0.5 \citep{hillenbrand_1998, da_rio_2014}.

Some authors \citep[e.g.\, ][]{da_rio_2017, kim_2019} claim that the velocity dispersion of
the stars of the ONC indicate that the cluster is in a virialized state, or
slightly supervirial. 
On the other hand, others \citep[e.g.\ ][]{  jones_1988, furesz_2008,
tobin_2009, da_rio_2014, stutz_2018, theissen_2022} claim that the system is not yet
virialized, and is either in expansion \citep{jones_1988, swiggum_2021}, or the
dynamics are still dominated by the gas \citep{furesz_2008, tobin_2009, stutz_2018}.
Numerical experiments \citep[e.g.\ ][]{kroupa_1999, kroupa_2001, scally_2005}
indicate that the best fit models for the ONC corresponds to models in
expansion, where the gas was  either absent \citep{kroupa_1999, scally_2005}, or
removed shortly after the beginning of the simulation \citep{kroupa_2001}, in
which they find that the ONC might evolve to a Pleiades-like cluster after
ejecting 2/3 of the initial stars. 
\citet{proszkow_2009} find that the observed properties of the ONC can be
explained by assuming a nonspherical cluster,  where stars have subvirial
velocities.

Observation of protostars and pre-main-sequence stars in the ISF show  that 
while protostars are located right on top of the ridgeline of the filament,
pre-main-sequence stars are symmetrically distributed around the filament
\citep{stutz_2016,beccari_2017, kainulainen_2017, stutz_2018}.
Other star forming regions, such as NGC1333, also show younger stars near the
gas filaments, while older stars are distributed more uniformly within the gas
cloud \citep{foster_2015,hacar_2017a}.
The radial velocity of the stars relative to the gas is also different for the
younger and older populations, where protostars have radial velocities close to
the velocity of the gas, with a low velocity dispersion.
Meanwhile, the pre-main-sequence stars have a larger radial velocity dispersion,
of the order of, or even larger than, the velocity dispersion of the gas
\citep{stutz_2016}.
A larger distance from the ridgeline of the filament, plus a larger velocity
relative to the gas, implies that the older stars have more kinetic energy than
the protostars.
The gas of the ISF presents undulations, not only in space, which gives the
filament its name, but also in velocity~\citep{gonzalez_2019}.
The regularity of these undulations suggests that the filament is being
subjected to strong transverse forces \citep{stutz_2016,stutz_2018a}.

To explain these observations, \citet{stutz_2016} proposed a scenario where the
gas of the ISF oscillates.
\citet{stutz_2016} named this scenario ``the Slingshot'', in which the movement
of the filament is injecting energy into the stellar system, increasing the
velocity and spread of the older stars.
In this scenario, the filament is moving;  inside the filament the gas starts
to collapse to form a protostar.
Since the protostar is forming from the filament, it has a low velocity
relative to the filament.
Once the protostar accretes enough mass, it decouples from the gas.
In this decoupled state, when the filament starts to decelerate, the star is
not able to stay in the ridgeline of the filament, and is ejected; that is, it is
not the stars who leave the filament, but the filament is leaving the stars
\citep{stutz_2016}.

\citet{stutz_2016} note that, at the scale of the filament as a whole, the
ratio of gravitational and magnetic energy indicates that the magnetic fields
are supercritical, while at scales of $\sim$1~pc, the fields are subcritical.
This balance between the magnetic and the gravitational field would allow the
ISF not only  to suffer violent periodic perturbations, which would create the
conditions needed to trigger the formation of clusters and eject stars, but 
also to survive said perturbations.

\citet{stutz_2018} assumed a spherically symmetric density distribution to
model the ONC stars.
Their best match was a Plummer model \citep{plummer_1911}, with scale radius
$R_\textrm{pl}=0.36$~pc, and a central density of 5755 \Msun~pc$^{-3}$.
They determined that the gravity of the gas dominates over the gravity of the
stars  at all radii, with the exception of $r = 0.36$~pc, which corresponds to
the scale radius of the ONC.
At that point, the gravity of the stars has the same magnitude as the gravity
of the gas.
They also estimate the crossing time of the cluster, finding a value of
$\sim0.55$~Myr, very similar to the estimate of the gas filament motion
\citep[0.6~Myr,][]{stutz_2016}.
The fact that the scale length of the cluster has the same value as the
distance at which the cluster gravity and the gas gravity have equal magnitude,
suggests that the filament regulates the development of the structural
parameters of the cluster.

\citet{dominik_2018} show that, when taking into consideration the effects of
a magnetic field on the gas,  the system evolves to a state where
gravitational, rotational and magnetic energy are comparable.
Under these conditions, any perturbation in the filament will give rise to
periodic oscillations.
When applying these results to the ISF, they obtain an oscillation period of
2.9~Myr, comparable to the timescales estimated by \citet{stutz_2016} and
\citet{boekholt_2017}.

The slingshot scenario was tested by \citet{boekholt_2017}.
They test if an oscillating gas filament is able to reproduce some of the
observed properties predicted by the slingshot.
They represent the gas filament by using a cylindrically symmetric, analytical
density profile with a polytrope-like softening.
This filament is constantly accelerating, with the position of the central
part of the filament moving along the $x$-axis following a sinusoidal function.
Embedded in the oscillating potential, they place a string of point mass
particles, with small deviation from the centre of the filament and zero
initial velocity.
As the filament moves, so do the particles in the string of stars.
They found that an initially narrow distribution of stars can be dynamically
broadened by the oscillation of the filament.
The fraction of particles ejected by the filament at each oscillation depends
on the maximum acceleration of the motion.
When the fraction of particles is equal on both sides of the filament, the
particles have a non zero velocity with respect to the filament.

In an effort to explain the ONC, \citet{kroupa_2018} also provide an
alternate idea for how the ONC formed. In their scenario, very young
clusters with masses in the range 300-2000 \Msun are able to
suppress stellar formation due to the presence of O stars which ionise
the gas and reduce the gas inflow, eject the ionising stars via
dynamical interactions, and then restart the star formation phase,
producing populations with different ages. With the expulsion of the
gas, the cluster can also expand, explaining the different spatial
distributions of these populations.
Another explanation for the extended distribution of the stars in the ISF
comes from three body interactions between stars in the ONC.
Three body systems are unstable, and sooner or later, one of the bodies
will be ejected \citep{valtonen_2006}.
\citet{reipurth_2010} use three body interactions with a background
potential to study the evolution of primordial binaries within star forming
clouds.
They subtract mass from the background potential to simulate the destruction of
cloud cores.
While the triple system manage to eject stars outside the cloud core, to
distances comparable to the radial extent of the ONC, the escaping stars do
not have the high velocities observed in the ISF.

Following \citet{boekholt_2017}, in this work we continue the exploration of
the effects of the Slingshot scenario on embedded star clusters like the ONC.
In this work we replace the string of stars with a spherical cluster of stars,
representing the ONC, and study the effects of the filament, both static and in
oscillation, on the dynamical evolution, and possible destruction, of the
cluster.

This paper is structured as follows.
In Section~\ref{sec:setup} we show the filament model used, the methods that we
employ to generate the star cluster, their initial masses and radii, and the
software used to run the simulations.
Section~\ref{sec:estatico} covers the case for a static filament.
In Section~\ref{sec:res} we explore the effects of the oscillating filament on
clusters of different masses and radii.
Finally, in Section~\ref{sec:conc} we present a summary of the results and our
conclusions.

\section{Method}\label{sec:setup}

In this section we describe the code used for the simulations
(Section~\ref{subsec:amuse}). 
Then we continue with a description of the model we use for the filament
(Section~\ref{subsec:fil}) and the star cluster (Section~\ref{subsec:cl}).  
For the filament we assume a softened power law cylindrical density profile and
sinusoidal oscillations.  
For the star cluster we assume a spherical Plummer density profile, and vary
the mass and radius. 
We finish this section with a description of the initial conditions and
parameters used in the simulations (Section~\ref{sec:init}).

\subsection{\amuse\ and \bridge}\label{subsec:amuse}

Our experiment requires a numerical framework capable of solving the N-body
(where $N=1000$) cluster problem self-consistently with a time-dependent
background potential (the filament, Section~\ref{subsec:fil}).
The Astrophysical Multi-purpose Software Environment
\citep[\amuse;][]{mcmillan_2012, pelupessy_2013, portegies_2013, amuse_2018}
allows us to accomplish this.
\amuse~is the astrophysical implementation of \textsc{muse}
\citep{portegies_2009}, a software framework that has the capability to combine
computational tools for different physical domains, allowing for
self-consistent  multi-physics simulations.
For this project, we use \phf~\citep{mcmillan_2012,zwart_2014}, a 4th order
Hermite predictor-corrector N-body code, written in \textsc{C++}, to update the
position and velocities of the particles.
\phf\ can be compiled with GPU support, which we use to speed up our
simulations.

We also need a way to account for the effects of the gas filament.
The filament is represented as a analytical background potential
(Section~\ref{subsec:fil}).
Instead of adding the background potential directly in the source code of \phf,
we use the \bridge~method \citep{fujii_2007}.
\bridge~provides a way to couple different codes to obtain a self-consistent
simulation. 
In our case, \bridge~couples \phf~with the background potential, so that the
particles in the cluster will move under the gravity of the filament.
To avoid numerical effects, the time-step of the simulation must be chosen
carefully.
A small value will give an increased precision when calculating the sum of the 
forces acting on a particle, but at the cost of increased CPU time.
The \bridge~timestep, effectively the timestep of the \mbox{cluster-filament}
system, is set to 100 yr.
This timestep is equivalent to $6\times10^{-3} t_\mathrm{cross,min}$, where
$t_\mathrm{cross,min}$ is the smallest crossing time between the models of
Table~\ref{tab:init}.

The total time of the simulation, $T_\mathrm{sim}$, is two times the 
oscillation period, or 2~Myr, whichever is larger.
This allows us to study the effects of at least one full oscillation cycle on
the cluster.
For simulations regarding the static filament case, the system is let to evolve
for a total of 2 Myr. 
To study the evolution of the system, a series of snapshots are taken at regular
intervals.
Each snapshot is taken every $T_\mathrm{sim}/500$, for a total of 500 
snapshots per simulation.
Numerical studies of fragmentation in turbulent gas clouds
\citep[e.g.][]{seifried_2015, clarke_2017} show that filaments with line mass
density comparable to the ISF fragment and form stars on timescales of a few
tenths of Myr.
This will change the density profile of the gas filament, reducing the effect
of the Slingshot on the cluster.
Although we run our simulations for times of up to 10~Myr, much longer than the
fragmentation timescale, we are interested in exploring the oscillation
parameter space in the idealized case where the filament exists for several
Myr, which include periods that are longer than the lifetime of real filaments.
As we will se below (Sections~\ref{sec:heal} to \ref{sec:ejected}), the effects
of the moving filament on the cluster happen within the first 1/4 oscillation,
so this length of time will be enough to eject, or destroy, the young cluster.

\subsection{The filament}\label{subsec:fil}

Based on dust maps of \citet{stutz_2015} and the analysis published by
\citet{stutz_2016}, \citet{stutz_2018} calculated a mass density profile of the
ISF at the position of the ONC.
They show that the gas density and gravitational potential of the filament
within $0.05 < r < 8.5$~pc follows a power law profile with a power law index
of $\gamma = 0.225$.
Even though the potential is well behaved at $r=0$, the gravitational
acceleration and volume density diverge.
Since the density profile must flatten at some point, we follow the method
developed by \citet{boekholt_2017}, to model the filament as a cylindrically
symmetrical potential. 
This potential extends to infinity along the z axis, with the profile being
constant along said axis, and a radial dependence that follows a power law with
a polytrope-like softening:
\begin{eqnarray}
    \rho(r)_\mathrm{model} &=& \rho_0\left[1+\left(\frac{r}{D}\right)^2\right]^\frac{\gamma-2}{2}\\ \label{eq:lambda}
    \Lambda(r)_\mathrm{model} &=& \Lambda_0 \left\{ \left[ 1 + \left(\frac{r}{D}\right)^2 \right]^\frac{\gamma}{2} - 1 \right\}\label{eq:mass}\\
    g(r)_\mathrm{model} &=& 
    \begin{cases}
        -\frac{2G\Lambda(r)}{r}\hat{r} & r>0\\
        0\hat{r} & r=0
    \end{cases}\\
    r^2 &=& (x_\mathrm{fil} - x)^2 + (y_\mathrm{fil} - y)^2
\end{eqnarray}
where $\rho_0 = 7.6\times 10^{10}$~\Msun pc$^{-3}$ is the density at the
filament axis, $D$ is the softening radius, $\gamma$ is the power law of the
gas profile, and $\Lambda_0 = 2\pi\rho_0 D^2\gamma^{-1}=53.07$~\Msun pc$^{-1}$\
is the line mass density.
The $D$ parameter regulates the curvature of the model.
When $r \gg D$, our models converge to the power law observed by
\citet{stutz_2016}.
We adjust the value of the softening radius $D$ so that the difference between
the value given by our model (Equation~\ref{eq:lambda}) and the value of the
line mass density at 8.5~pc, observed by \citet{stutz_2016}, is less than
$10\%$.
We set the value of the softening parameter to $D=5\times10^{-6}$~pc.
While this value is small, on the order of 1~AU, a larger value would increase
the curvature of the profile within the region observed by \citet{stutz_2016}.
The quantities $x_\mathrm{fil}$ and $y_\mathrm{fil}$ correspond to the position
of the ridgeline, and are defined below (Equations~\ref{eq:pos_x} and
\ref{eq:pos_y}).
These profiles are valid within the few inner parsecs from the filament centre.
The enclosed mass, given by the density profile, will grow without bounds,
reaching an infinite mass when integrated to infinity.

The observations from \citet{stutz_2016} extends to $\sim8.5$~pc and do not
show a cut-off.
Since most of our stellar interactions occur within that radius, the cut-off in
the density profile will be of limited importance in our result and it is not
included in the model.
The softening radius $D$, central density $\rho_0$ and power law of the gas
profile $\gamma$ have the same value for all simulations.

The slingshot scenario indicates that the ISF is a standing wave
\citep{stutz_2018a, gonzalez_2019}.
To mimic the motion of the filament, we use a sinusoidal function to determine
the position of the gas potential ridgeline:
\begin{eqnarray}
    x_\mathrm{fil}(t) &=& A\sin\left(\frac{2\pi}{P} t\right) \label{eq:pos_x}\\ 
    y_\mathrm{fil}(t)&=&0 \label{eq:pos_y}
\end{eqnarray}
where $A$ corresponds to the amplitude of the oscillation and $P$ is the period
of the oscillation.
From the values $A$ and $P$, we can also obtain the maximum velocity and
maximum acceleration of the filament:
\begin{eqnarray}
    v_\mathrm{max} &=& \frac{2\pi}{P}A\\
    a_\mathrm{max} &=& \left(\frac{2\pi}{P}\right)^2A
\end{eqnarray}

For small values of $A$, or large values of $P$, this model reduces to the
static filament model.
Even though the movement of the filament might be more complex in reality, this
function is the simplest model that can be analysed in detail.
The values of the parameters $A$ and $P$ are shown in Table~\ref{tab:init}.

\begin{table*}
    \caption{ 
        Parameters of the Plummer models (Plummer radius, Plummer mass and
        relaxation time of both an isolated Plummer sphere and a Plummer sphere
        embedded in the filament), plus oscillation parameters range used in
        the simulations and mass ratio between the filament and the cluster, at
        the Plummer radius.
        From these model sets, model D (highlighted with a star) uses the mass and
        radius of the ONC derived from \citet{stutz_2018}.
        We use 10 values for the oscillation amplitude, and 10 values for the
        oscillation period, within the respective range, for a total of 100
        filaments per model.
    } 
    \label{tab:init}
    \begin{tabular}{l l r l l c c c}
        \hline
        Model & $\Rpl$ &  $\Mpl$ &$t_\mathrm{relax,nogas}$& $t_\mathrm{relax,gas}$& Amplitude range & Period
        range & $\Lambda(\Rpl)/M_\mathrm{cl}(\Rpl)$\\
              &   [pc] & [\Msun] & [Myr] &[Myr] & [pc] & [Myr]&   \\
        \hline
        A &  0.1  & 250 & 2.54 & 1.73 & 0.05 - 5.0 & 0.5 - 5.0 & 4.97  \\
        B &  0.1  & 500 & 1.79 & 1.42 & 0.25 - 5.0 & 0.5 - 5.0 & 2.48  \\
        C &  0.1  & 1000& 1.26 & 1.21 & 0.5  - 4.0 & 0.5 - 5.0 & 1.24  \\ 
D$^\star$ &  0.36 & 1124& 8.17 & 5.41 & 1.0  - 2.5 & 0.5 - 3.5 & 1.52  \\ 
        \hline
    \end{tabular}
    \newline
\end{table*}

\subsection{The Cluster}\label{subsec:cl}

The star cluster is represented by a Plummer sphere \citep{plummer_1911} with
1000, equal mass ($m_\mathrm{particle} = \Mpl/1000$), particles.

We use different values for the Plummer radius $R_\mathrm{pl}$ and the Plummer
mass $M_\mathrm{pl}$ to study the way different clusters react to the motion of
the filament (Table~\ref{tab:init}).
For all models, the mass of the filament dominates over the mass of the cluster
at a distance of $r = \Rpl$ (Table~\ref{tab:init}).
Furthermore, we use a fixed particle number to prevent the low mass models from
having too few particles, and for the high mass models to use too much computer
time.

The addition of an external potential will affect the dynamics of the cluster.
In isolation, a Plummer model will keep its spherically symmetric distribution
\citep{gal_dyn}, but our cylindrical potential will prevent this.
Due to the symmetry of the potential, the gas will exert a force only in
the direction perpendicular to the filament, i.\ e.\ in the $x-y$ plane.
To prevent a collapse of the cluster in the  direction perpendicular to the
filament, we increase the $x-y$ velocities of the particles, adding enough
energy to the stars so that the combined system is stable and not collapsing.
The magnitude of the augmented velocity for a given particle is drawn from
the distribution function of a Plummer sphere \citep{gal_dyn}.
Instead of using the escape velocity from an isolated Plummer sphere at the
position of the particle, we use difference between the potential of the
filament-cluster system at the position of the star and at a distance of
$5\Rpl$, effectively increasing the maximum velocity that the stars can
have. 
This way, particles will have enough energy to reach, at most, a distance of
$5\Rpl$; any star that has moved beyond this limit must have had its
energy increased via dynamical interactions with the filament, other stars of
the cluster, or both.
The new velocity $v_\mathrm{mod}$ is assigned to the particle, keeping the
original $z$ component of the velocity, and modifying the $x$ and $y$
components of the velocity so that the magnitude of the new velocity is
$v_\mathrm{mod}$.

\subsection {Initial Conditions}\label{sec:init}

To explore the effects of an oscillating gas potential on the dynamics of an 
embedded star cluster, we use clusters with different Plummer radii and total
masses.  
A total of 15 combinations of Plummer radius and total masses  for the clusters
(Table~\ref{tab:init}) were used.
In all cases, the initial position of the centre of mass of the cluster
corresponds with the ridgeline of the filament.
The clusters start with zero initial velocity.
Before the filament begins to move, the cluster is left to evolve inside a
static filament for a period equivalent to 3 times the crossing time of a
Plummer sphere with the same total mass and Plummer radius than the model.

Sets A,B,C and D explore the effects of the oscillating potential on the
centre, by using filaments with different amplitude and period
(Section~\ref{sec:res}).
Each set uses ten, equally spaced, values for the amplitude and period of the 
oscillation, for a total of 100 filaments per cluster model.

\section{Static Filament}\label{sec:estatico}

\begin{figure}
    \centering
    \includegraphics{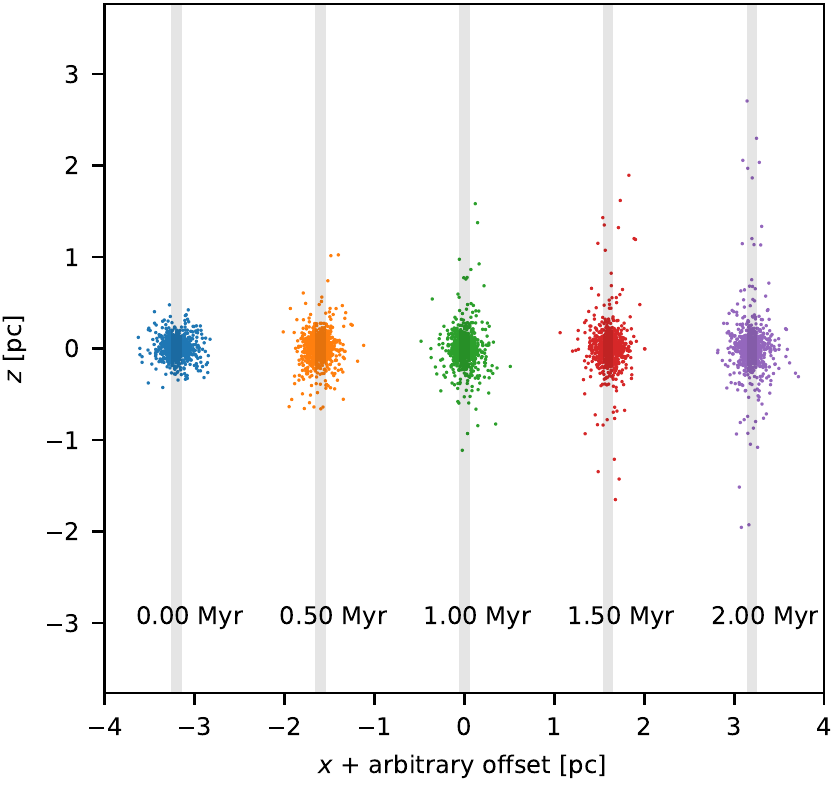}
    \caption{Snapshots of a cluster with $\Rpl=0.1$~pc, $\Mpl=500$\Msun \,inside
    a static filament.
    Due to the symmetry of the potential, the particles of the cluster are free
    to move along the filament, following ``corkscrew'' orbits around the ridgeline.
    As a result, the cluster evolves to a non-spherical equilibrium shape.
    }\label{fig:estatico_snap}
\end{figure}

Setting the amplitude of the oscillation to $A=0$~pc, we obtain the static
filament case.
Although the ISF is not static \citep{stutz_2016}, we consider first the case
with $A=0$~pc to demonstrate the stability of the cluster, and to provide a
benchmark for the evolution of the cluster without gas motions.
We study the evolution of a cluster in a static filament by using models with
different Plummer masses (100~\Msun, 500~\Msun~and 2000~\Msun) and Plummer
radius of 0.1~pc (Table~\ref{tab:estatico}).
The gas mass enclosed within the Plummer radius is, according to
Equation~\ref{eq:mass} is $\Lambda(0.1~\mathrm{pc})_\mathrm{model} = 440$~\Msun.
We evolve the clusters inside the filament for 2 Myr.

Due to the symmetry of the potential, the gas filament only produces forces
parallel to the $x-y$ plane, therefore, the particles are free to move in the
$z$ direction, with ``corkscrew'' orbits around  the filament.
These ``corkscrew'' orbits may have observational effects on proper motion and
radial velocity data of young stars in gas filaments. 
This has the effect of extending the particle distribution
(Figure~\ref{fig:estatico_snap}), transforming the initially spherical cluster
to a system elongated in the direction parallel to the filament. 
Nonetheless, the cluster is still attracting these particles, so they will
eventually stop moving away from the cluster and will fall back. 
Particles with $v > v_\mathrm{esc,cl}$, which are unbound from the cluster but
still bound to the filament, will keep moving along the filament without
falling back into the cluster.

\begin{table}
    \caption{Plummer models used for the static filament. 
    }\label{tab:estatico}
    \begin{tabular}{r c c}
        \hline
        $\Mpl$ & $\Rpl$ & Simulation time \\ 
       \, [\Msun] &  [pc]  & [Myr]           \\
        \hline
        100    &   0.1  & 2.0 \\
        500    &   0.1  & 2.0 \\
        2000   &   0.1  & 2.0 \\
        \hline
    \end{tabular}
\end{table}

\begin{figure}
    \centering
    \includegraphics{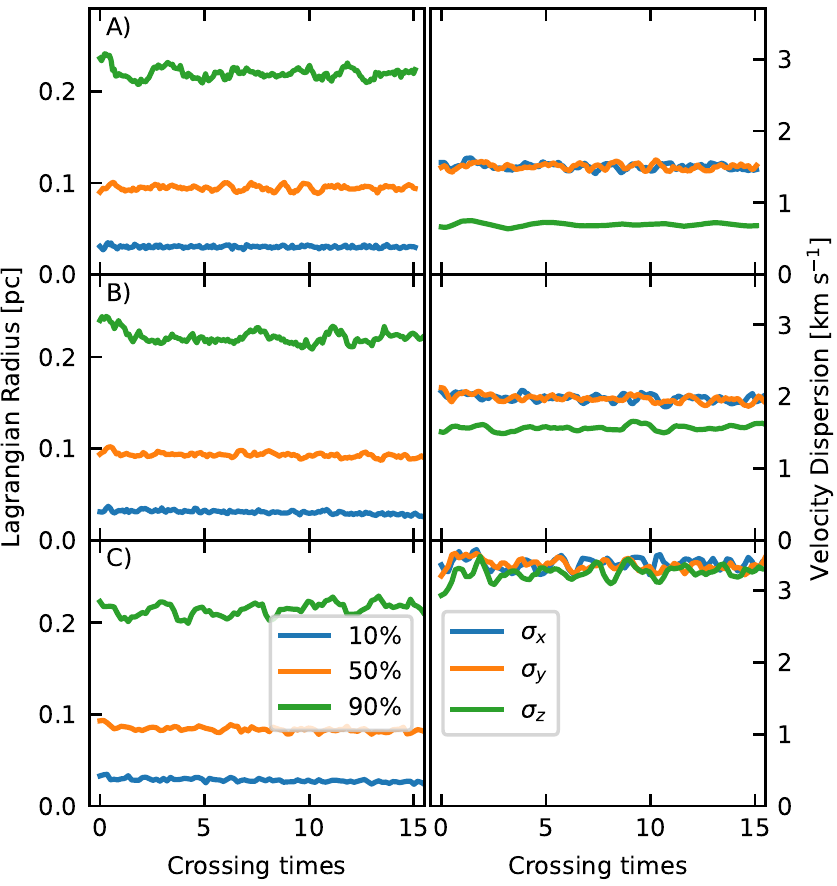}
    \caption{ Lagrangian radii (left) and Velocity dispersion (right)  for
    clusters inside a static filament.
    All panels correspond to clusters with the same Plummer radius ($\Rpl =
    0.1$~pc), but different masses: A) 100~\Msun, B)500~\Msun, and C) 2000~\Msun.
    For panels in the left column, the colours indicate the amount of mass
    enclosed within the respective radius: blue lines for 10\% of the total
    mass, orange dashes for 50\%, and green dash-dotted for 90\%.
    With increased $x-y$ velocity of the particles, the Lagrangian radii of
    each cluster remain constant.
    This velocity increase also produces an anisotropic velocity dispersion,
    with a lower value in the direction parallel to the filament ($z$
    direction, green line). 
    In more massive models, the difference between $\sigma_z$ and
    $\sigma_{x,y}$ decreases, as the dynamics of more massive models are less
    dominated by the gas potential, as expected.
    }\label{fig:2_plot}
\end{figure}

The stability of the cluster can be seen by plotting the Lagrangian radii of
the cluster (Figure~\ref{fig:2_plot}, left panels).
They show almost no change during the 2~Myr that the simulation last.
The velocity adjustment (Section~\ref{subsec:cl}) will also increase the
velocity dispersion in the $x$ and $y$ axis, while also preserving the original
value in the $z$ direction.
Figure~\ref{fig:2_plot}, right panel, shows the difference between the velocity
dispersion in these two directions.
For less dense clusters, the ratio $\sigma_x/\sigma_z$ is larger than for more
massive clusters.
This difference tells us that clusters with shallower potentials need a larger
velocity boost to prevent collapse due to the filament.
On the other hand, the cluster  with $\Mpl = 2000$\Msun\, has similar velocity
dispersion in all axes, meaning that a more dense object does not need a
velocity boost to prevent contraction.

\section{Oscillating Filament}\label{sec:res}

Motivated by the observational evidence outline above
(Section~\ref{sec:intro}), we aim to study the effect of a time-dependent
potential on the cluster structure, specifically that of an ``oscillating
filament''. 
One of the effects of the oscillation of the filament on the cluster will be
the loss of mass from the cluster, so we try to quantify the degree of mass
loss and particle ejection.
The models that we use for the filament (Section~\ref{subsec:fil}) do not
include a cutoff radius for the gas mass and, therefore, there is no
well-defined escape velocity for the particles.
Instead, we use a practical criterion that consists of particle displacement
from the cluster centre or potential ridgeline.
We define a particle as escaped from the filament if $r > 5\Rpl$ from the
centre of the filament, or escaped from the cluster if $r > 5\Rpl$ from the
centre of density~\citep{cartesano_1985} of the cluster.
This distance is chosen since no particles are located beyond that radius at
time $t=0$.
Once the filament has completed one full oscillation, we count the number of
particles that are inside these limits.

\begin{figure}
    \centering
    \includegraphics{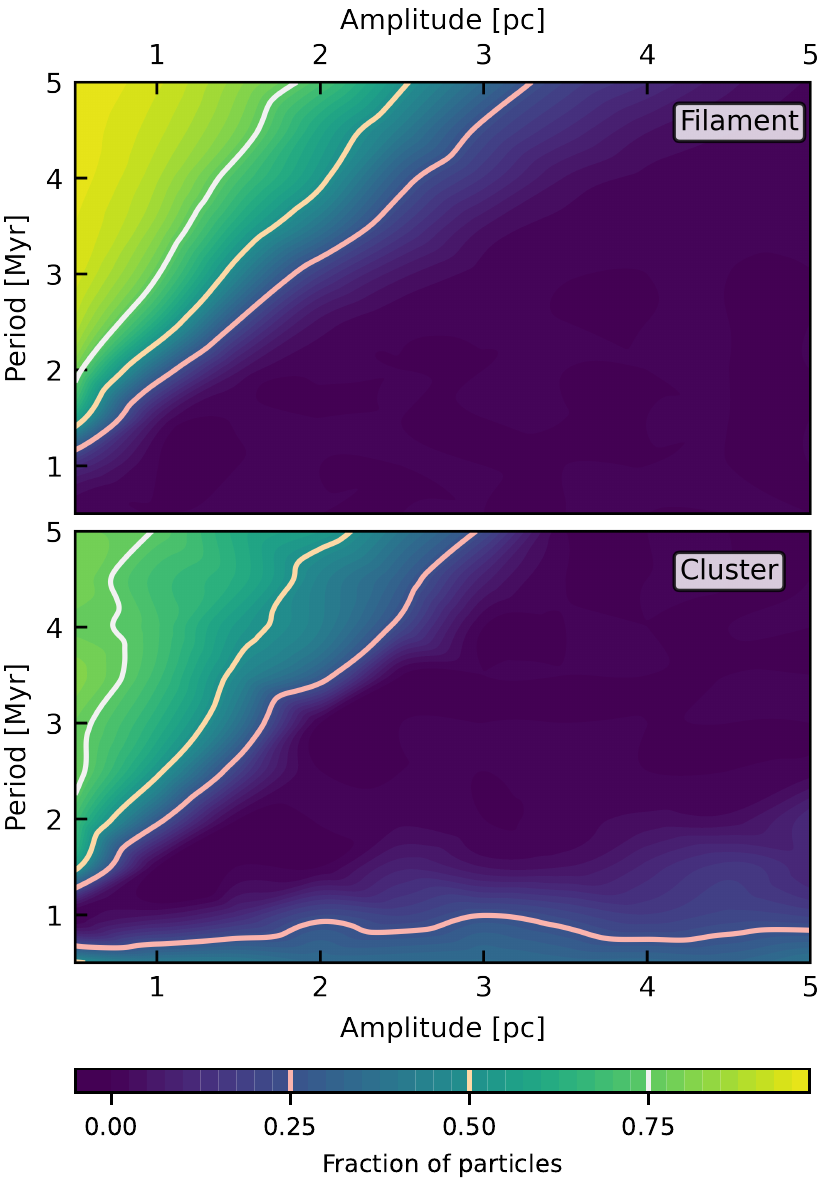}
    \caption{Fraction of particles left in the filament (top) and in the
    cluster (bottom) for different combinations of oscillation amplitude and
    oscillation period for Model A ($\Rpl=0.1$~pc, $\Mpl=250$~\Msun).
    The color code indicates the fraction of particles after one full
    oscillation, with contours for 25\%, 50\% and 75\% of the initial number of
    particles.
    The fraction in the filament decreases smoothly as the maximum velocity of
    the filament increases.
    A similar trend is observed for the fraction of particles in the cluster,
    with a increase in the number of particles if the filament is moving with a
    large velocity.
    }\label{fig:pera_fracs}
\end{figure}

Figure~\ref{fig:pera_fracs} shows the retained particle fractions of a cluster
with a total mass of $\Mpl=250$~\Msun\ and Plummer radius $\Rpl = 0.1$~pc.
These values correspond to 100 realizations of Model A in Table~\ref{tab:init},
each inside a filament with different values of oscillation period and
oscillation amplitude.
Top panel shows the fraction of particles left inside the filament at the end
of one oscillation, while the bottom panel shows the fraction of particles in
the cluster.
In both panels, the color code indicates the fraction of particles left in the
respective structure at the end of the first oscillation, with contours for
75\%, 50\% and 25\% of the initial number of particles.
For small amplitudes and large periods, most of the particles remain inside the
filament (top) and the cluster (bottom).
As the maximum velocity of the filament increases, a larger fraction of
particles are ejected from the system, up to the point where almost all the
particles have left the filament, and there is no longer a cluster of stars,
only streams of particles moving around the gas potential.
Filaments with large maximum velocities will cross the cluster quickly, and the
potential will not have enough time to accelerate the cluster.
Hence, the cluster remains close to its initial position.
With these quick passages, the filament cannot inject enough energy to
completely destroy the cluster, so the it will be able to keep a small,
but not zero, fraction of stars.

\subsection{Outcomes for the clusters}\label{subsec:4zone}

\begin{figure}
    \centering
    \includegraphics{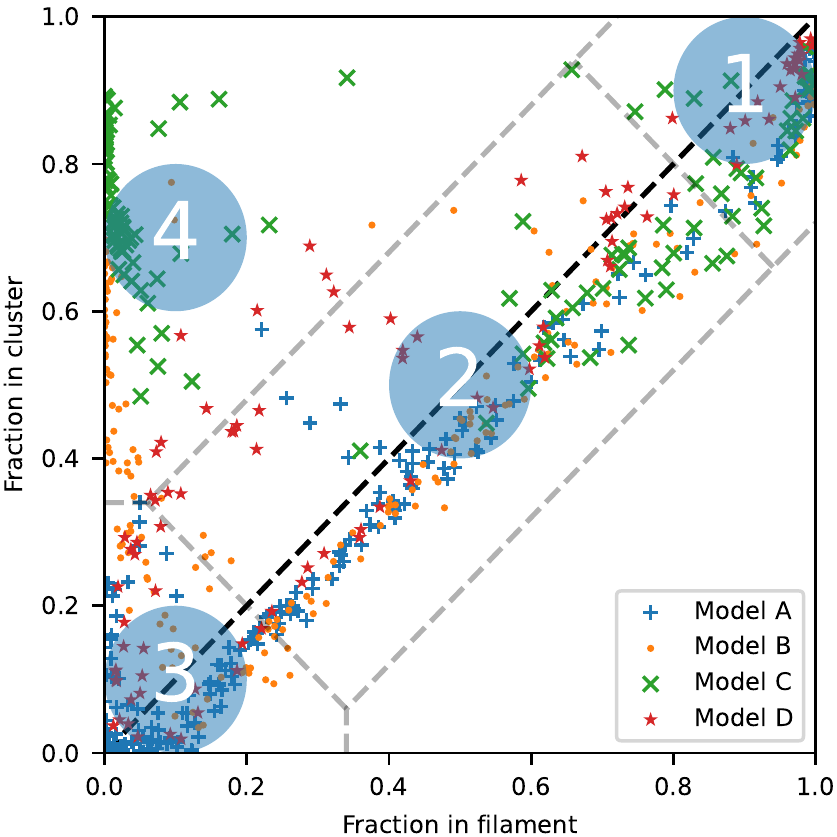}
    \caption{ Fraction of particles in the cluster versus the filament at the
    end of the first oscillation for models with fixed $\Rpl = 0.1$ (all models
    in sets A, B and C, plus model D, with $\Rpl=0.36$ pc, see
    Table~\ref{tab:init}).  
    Each point in the plot corresponds to a simulation with different values of
    the oscillation amplitude and oscillation period. 
    Each symbol represents a model, with blue plusses indicating simulations
    from Model A, orange dots for Model B, green crosses for Model C and red
    stars for Model D.
    We classify the remnant by its position in the fraction-fraction plot
    into one of four categories: 1) Filament Associated clusters, 2) Transition
    clusters, 3) Destroyed clusters, or 4) Ejected clusters (see main text for
    details on each category). 
    }\label{fig:zonesR01}
\end{figure}

For each pair of amplitude and period, we obtain objects with different
fractions of particles inside the filament and inside the cluster.
Figure~\ref{fig:zonesR01} shows the position of the resulting object in this
outcome space, where each point represents a different combination of
oscillation parameters.
The different symbols indicate the results for models A (blue plusses), B
(orange dots), C (green crosses), and D (red stars).
The masses and radii of these models are shown in Table~\ref{tab:init}.
The black dashed line represents a 1:1 ratio between the fraction of particles
left in the filament and in the cluster.
After one oscillation, most of the remnants are located near this reference
line.
For models near this reference line, this can be explained by the fact that when
a particle leaves the filament, it will also be ejected from the cluster, so
the fraction of particles in the cluster will follow closely the fraction in
the filament.
However, the opposite is not necessarily true: a particle can leave the cluster
by moving along the filament, in which case said particle will be counted as
``in the filament'' but not as ``in the cluster'', so the fraction in the
filament tends to be slightly larger than the fraction in the cluster.
Along this sequence we can identify three different outcomes for the cluster,
plus a fourth type of object located outside the 1:1 line, this fourth type
maintains most of the particles inside the cluster, and almost none inside the
filament, indicating that the cluster has left the central part of the gas
column.
We names these types as 1)~Filament Associated clusters; 2)~transition
clusters; 3)~destroyed clusters; and 4)~ejected clusters.

\begin{figure}
    \centering
    \includegraphics{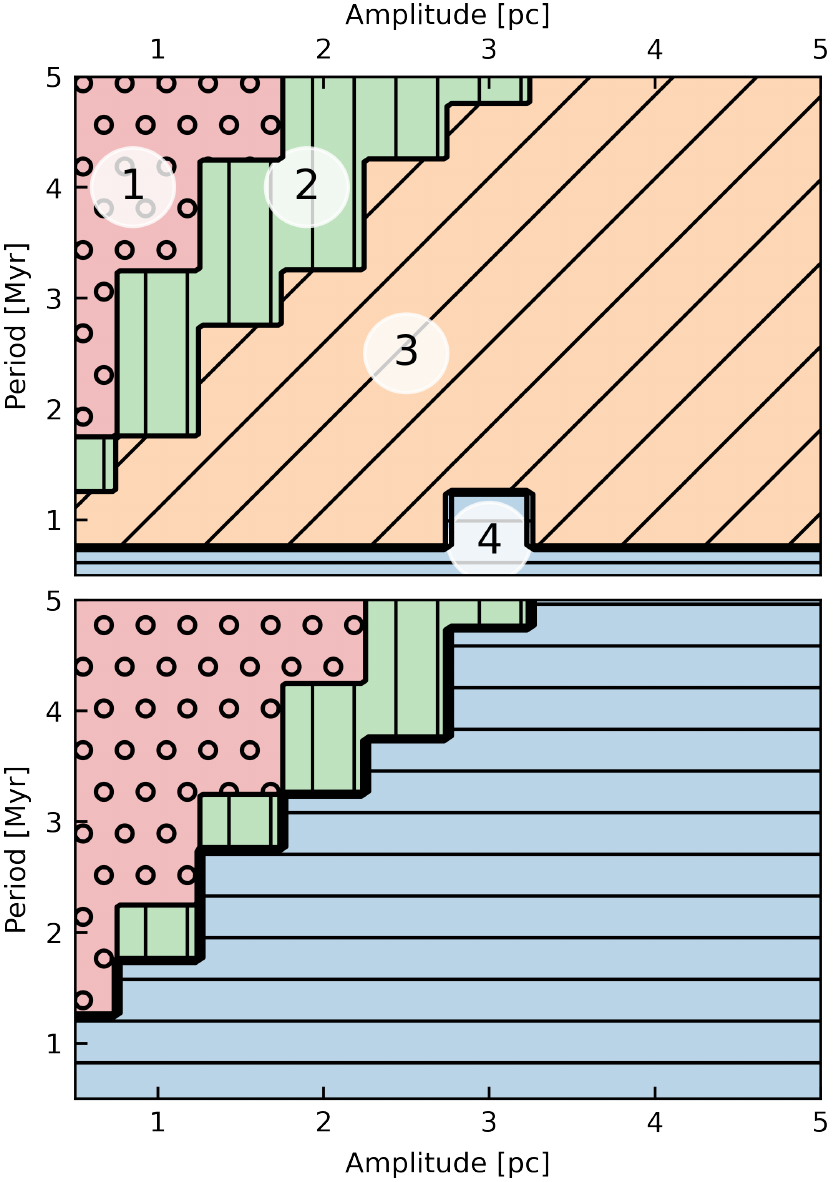}
    \caption{Classification of the remnants as function of the oscillation
    amplitude and period for models A (top) and C (bottom).  
    The coloured regions and the numbers correspond to the numbered zones from
    Figure~\ref{fig:zonesR01}, with the red circles indicating the Filament
    Associated clusters, green vertical bars for the transition clusters,
    orange diagonal bars for the destroyed clusters, and blue horizontal bars
    for the ejected clusters.
    Note that model C does not have destroyed clusters for any combinations of
    oscillation period and amplitude.
    }\label{fig:zonas_colores}
\end{figure}

Not all models generate remnants in all regions.
For example, the set of $\Rpl$ = 0.1~pc and $\Mpl$ = 1000~\Msun~models (set C
in Table~\ref{tab:init}) does not have destroyed clusters, either they lose a
small fraction of stars or eject the whole cluster, as can be seen by the
absence  of green crosses in the region marked by the number three in
Figure~\ref{fig:zonesR01}, which corresponds to the destroyed clusters.
This also can be seen by plotting which pairs of amplitude and period from
Figure~\ref{fig:pera_fracs} give rise to remnants belonging to each of the four
regions.
Figure~\ref{fig:zonas_colores} is such example for models A (top) and C
(bottom).
Each of the regions is represented by a different colour, following the
numbers used in Figure~\ref{fig:zonesR01}.
Filaments with small amplitude oscillations and large periods are the ones that
generate clusters for region 1, as mentioned above.
As we use a filament with larger maximum velocity, the resulting remnant will
be placed along the 1:1 track of Figure~\ref{fig:zonesR01}, generating objects in
the transition region (region 2), then destroyed clusters (region 3), and
finally generating ejected clusters (region 4).

If the filament is moving fast enough to eject the cluster, once the cluster
moves outside the central part of the filament, its particles will still have
the extra kinetic energy (see Section~\ref{sec:setup}) but without the gas
potential, the cluster will lose most of its stars until it is dissolved.
On the other hand, for more massive clusters, the relative adjustment to the
velocities that is required to maintain equilibrium in the filament is small
compared to the velocities that the stars would have in the absence of the
filament potential (see e.g.\ Figure~\ref{fig:2_plot}).
Hence, upon ejection from the filament, the stars remain, for the most part,
bound to the cluster.

\subsection{Filament Associated Clusters}\label{sec:heal}

\begin{figure}
    \centering
    \includegraphics{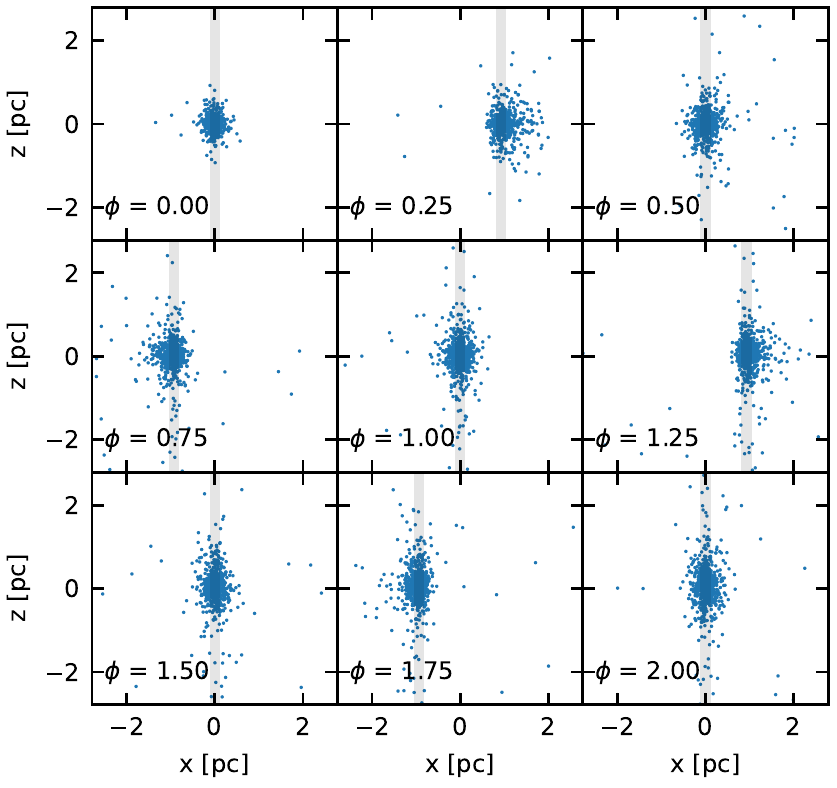}
    \caption{Time evolution of the stars (blue dots) in an
    oscillating gas filament (grey vertical line).
    The cluster shown here corresponds to model A ($\Rpl$=0.1~pc,
    $\Mpl$=250\Msun), inside a filament that oscillates with period P=3.5~Myr
    and amplitude A = 0.93~pc.
    The slow movement of the filament is not enough to eject a significant
    fraction of particles from the cluster, and so we classify this cluster in
    the ``Filament Associated Cluster'' category.
    The label indicates the oscillation phase, defined as $\phi=t/P$, where
    $\phi=1$ is the end of the first oscillation.
    }\label{fig:sano_snap}
\end{figure}

Filaments with low amplitudes and large periods, which translates to low
maximum velocities and acceleration, give rise to clusters in the ``Filament
Associated Cluster'' region.
They evolve in a similar way to clusters inside static filaments
(Section~\ref{sec:estatico}), with relatively constant velocity dispersion,
Lagrangian radii, and elongation along the filament.
Figure~\ref{fig:sano_snap} shows nine snapshots of model A ($\Rpl$=0.1~pc,
$\Mpl$=250\Msun), inside a filament oscillating with period P=3.5~Myr and
amplitude A=0.93~pc.
These snapshots cover a total of two full oscillations of the filament.
The cluster always stays close to the filament.

\begin{figure}
    \centering
    \includegraphics{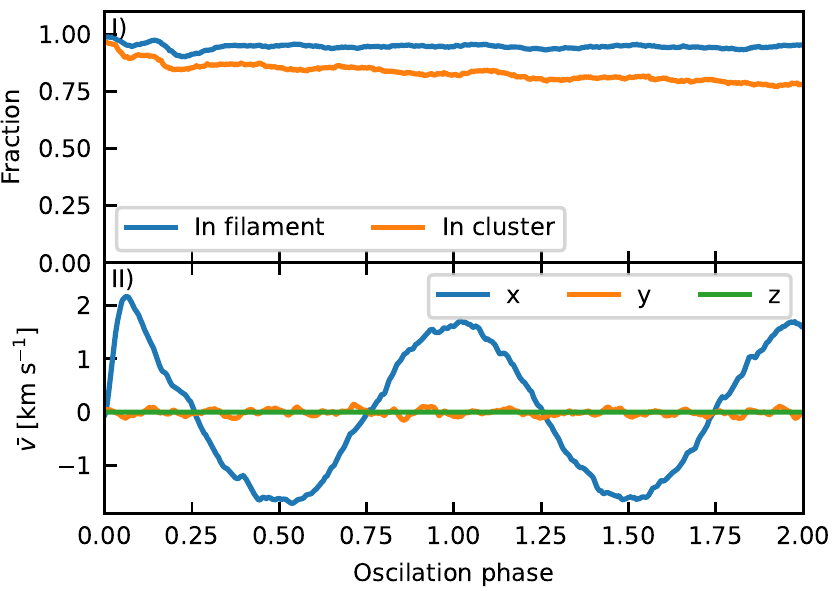}
    \caption{Time evolution of fraction of particles inside the cluster and
    inside the filament, and mean velocity of the particles for the simulation
    shown in Figure~\ref{fig:sano_snap}, which belongs to the ``Filament
    Associated'' clusters category.
    The top panel shows the fraction of stars inside the filament (blue line)
    and inside the cluster (orange line), as function of the oscillation phase.
    Second panel shows the mean velocity of the particle system for the three
    spatial directions (blue for $x$, orange for $y$, and green for $z$).
    The $x$ axis, ``Oscillation phase'' corresponds to the quantity $t/P$,
    where $P$ is the period of the oscillating filament.
    }\label{fig:sano_plots}
\end{figure}

At the beginning of the simulation, the cluster has zero mean velocity.
As soon as the filament moves, the cluster starts to fall towards the centre of
the filament.
Due to the random motions of the particles, a fraction of these particles will
have velocities with the opposite direction to the movement of the filament.
The sum of these two effects, the cluster falling towards the filament, moving
in the positive $x$ direction, and the particles moving in the negative
direction, is the cause of the initial ejection of particles, which can be seen
in Figure~\ref{fig:sano_plots}, top panel.
Even though the motion of the filament in the ``Filament Associated Cluster''
region cannot eject a large fraction of particles on its own, as reflected in
the almost constant value of the fraction of particles in the filament of
Figure~\ref{fig:sano_plots}, still some mass is lost due to the previously
mentioned mechanism.
Since the cluster is inside the filament, particles leave the cluster by moving
in ``corkscrew'' orbits around the centre of the gas potential.
This explains the downwards trend in the number of particles in the cluster,
but constant fraction in the filament: particles leave the cluster, but they
are unable to escape the filament.

The second panel in Figure~\ref{fig:sano_plots} shows the mean velocity of the
cluster for the three spatial axes.
The cluster starts with zero velocity, but it is attracted by the moving
potential, quickly falling back into the ridgeline of the filament.
It is during this moment when most of the ejections that the cluster
experiences take place.
After that initial acceleration phase, the cluster will move at the same
velocity as the filament for the rest of the simulation.

\subsection{Transition Clusters}\label{sec:trans}

\begin{figure}
    \centering
    \includegraphics{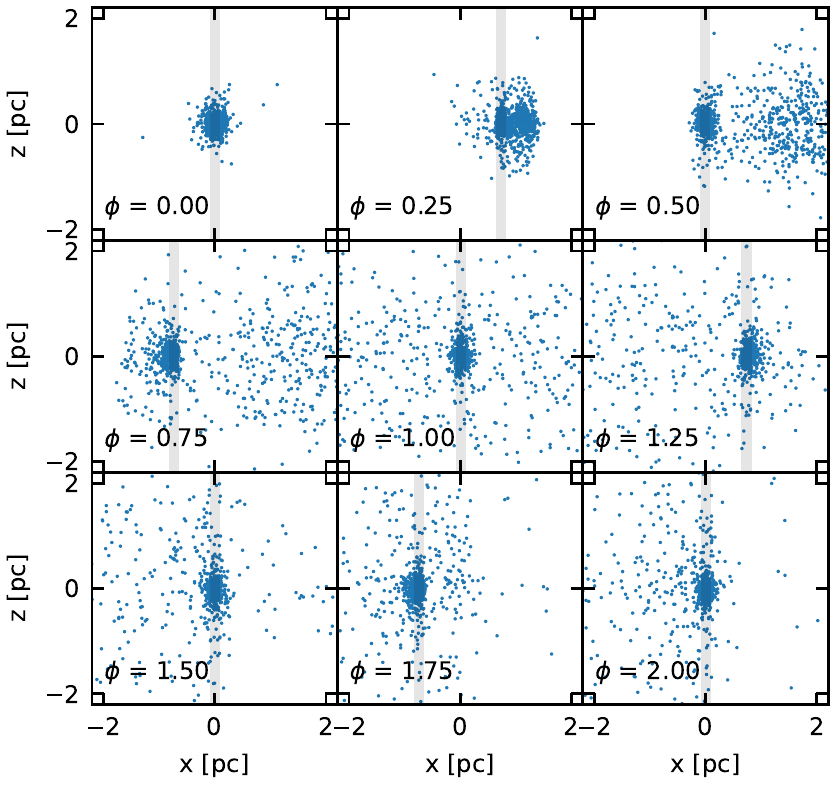}
    \caption{Time evolution of the stars (blue dots) in an oscillating gas
    filament (grey vertical line).
    The cluster shown here corresponds to model B ($\Rpl$=0.1~pc,
    $\Mpl$=500\Msun), inside a filament that oscillates with period P=1.73~Myr
    and amplitude A = 0.74~pc.
    Compared with the previous case (Section~\ref{sec:heal}), this model loses
    a larger fraction of particles at the first quarter of oscillation
    ($\phi$~=~0.25), but still retaining more than 20\% of the initial number
    of particles and, therefore, we classify it in the ``transition'' category
    (region 2 in Figure~\ref{fig:zonesR01}).
    }\label{fig:trans_snap}
\end{figure}

Increasing the maximum velocity of the filament, either by increasing the
oscillation amplitude, or by reducing the oscillation period, will increase the
number of particles lost by the filament and by the cluster.
This becomes evident after inspecting snapshots of this type of remnant.
Figure~\ref{fig:trans_snap} shows one example of a filament that produces a
remnant in the transition region (region 2 from Figure~\ref{fig:zonesR01}),
where we can see that a larger number of particles are ejected at $\phi=0.25$
(Figure~\ref{fig:trans_snap}, second panel) when we compare with the remnant
from Section~\ref{sec:heal}.
These snapshots correspond to a  filament with amplitude $A=0.74$~pc and period
$P=1.73$~Myr and a cluster with a total mass of 500~\Msun and a Plummer radius
of 0.1~pc (model B).

Clusters in this zone eject a considerable fraction of their initial
mass, but leave a remnant that stays inside of the filament. 
Since the cluster stays inside the filament, once a particles leaves the
filament, it also leaves the cluster.
The opposite is not true: a particle can move along the filament and reach a
distance larger than $5\Rpl$ from the centre of density of the cluster,
and once it crosses that boundary, it is no longer counted as belonging to the
cluster, but it is still inside the filament.
This is the reason why ``transition'' clusters are located under the diagonal
line in Figure~\ref{fig:zonesR01}, which acts as a reference for a $1:$1 mass
loss ratio, meaning that clusters in this zone have slightly more particles in
the filament than in the cluster.
Like clusters in a static filament (Section~\ref{sec:estatico}), or in the
``Filament Associated Cluster'' (Section~\ref{sec:heal}) category, they are
elongated in the direction of the filament (Figure~\ref{fig:trans_snap}, second
panel onwards).

\begin{figure}
    \centering
    \includegraphics{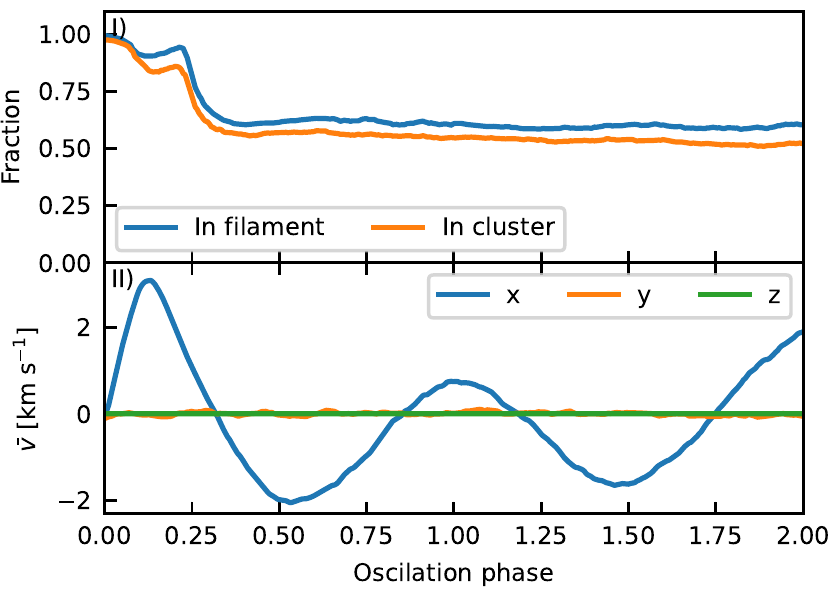}
    \caption{ Same as Figure~\ref{fig:sano_plots}, for Model B and the parameters
    from Figure~\ref{fig:trans_snap} (``transition'' cluster).
    The top panel shows the fraction of stars inside the filament (blue line)
    and inside the cluster (orange line), as function of the oscillation phase.
    Second panel shows the mean velocity of the stars for the three spatial
    axis.
    }\label{fig:trans_plots}
\end{figure}

\begin{figure}
    \centering
    \includegraphics{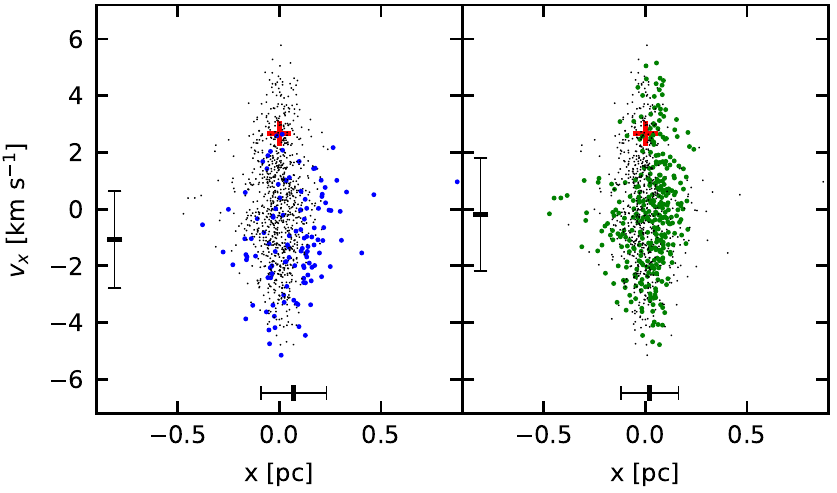}
    \caption{Phase space diagram showing the groups of stars ejected at the
    beginning (blue, left panel) and at turnaround (green, right panel) of the
    oscillation. 
    The position and velocity of the  filament represented with a red cross.
    The $x$ axis corresponds to the direction of the filament oscillation.
    This snapshot corresponds to a cluster from model B ($\Rpl$=0.1~pc,
    $\Mpl$=500\Msun), with the filament oscillating with $A=0.74$~pc,
    $P=1.73$~Myr.
    The bars at the left and bottom sides of the plot show the mean velocity
    and position of the highlighted particles relative to the filament, plus
    the dispersion on their values.
    In general, the particles ejected at the beginning are in front of the
    filament, and moving back towards the cluster.
}\label{fig:trans_groups}
\end{figure}

For remnants in the ``transition'' region, we can clearly distinguish two
phases of mass loss (Figure~\ref{fig:trans_plots}, first panel): the first
phase occurs right at the beginning of the simulation, when the cluster starts
to move towards the filament, and the second phase when the filament reaches
its maximum distance from its initial position, causing the cluster to stop
moving (to the right side in Figure~\ref{fig:trans_snap}) and ejecting the
second group of stars.
The first group of ejected stars corresponds to a small fraction of particles
located mainly in front of the filament.
These particles will attempt to fall into the cluster and into the filament,
moving in opposite direction to the filament, and so are left behind the main
bulk of stars (Figure~\ref{fig:trans_groups}, left panel).
The second group of ejected stars follow a similar process, but they stay
closer to the filament than the first group, and leave the cluster once the
filament starts to recede back to its initial position
(Figure~\ref{fig:trans_groups}, right panel).
For both groups, once the ejected stars reach and cross the filament, they are
moving too fast to be recaptured either by the filament or by the cluster.
After the ejection of the second group of stars, the fraction of particles in
the cluster remains constant during the remainder of the simulation, with only
a handful of particles being ejected each time the filament reaches its maximum
displacement (Figure~\ref{fig:trans_plots}, top panel).

\begin{figure}
    \centering
    \includegraphics{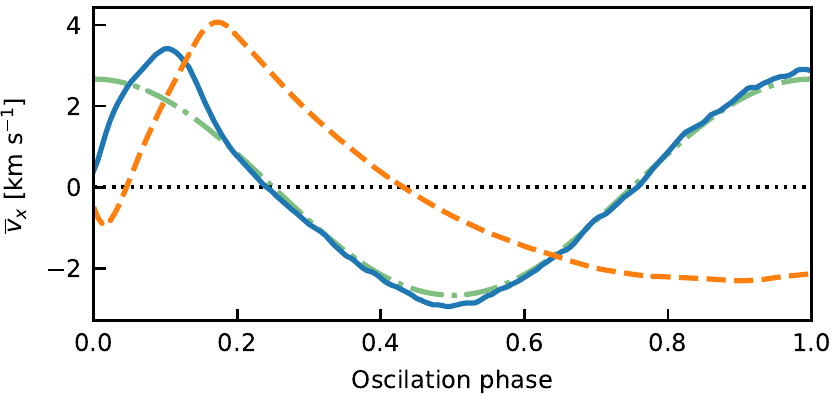}
    \caption{Mean velocity of the particles inside the filament (blue line) and
    of the ejected particles (orange dashed) for the simulation shown in
    Figure~\ref{fig:trans_snap}.
    The velocity of the filament is shown with the light green dash dotted line.
    As expected, the particles inside the filament move at the same velocity as
    the filament.
    On the other hand, the ejected particles have a larger maximum velocity
    than the particles inside the filament, which they reach at a later time.
    After that, the stirring of the filament causes the particles of this group
    to move in different directions, lowering their mean velocity.
    }\label{fig:trans_vmean}
\end{figure}

The motion of the filament drags the particles only in the $x$ direction.
This translates to almost zero mean velocity in the $y$ and $z$ direction, as
can be clearly seen in Figure~\ref{fig:trans_plots}, lower panel.
The cluster starts at rest, but as it accelerates due to the motion of the
filament, it reaches a maximum velocity of $\sim3.0$\kms, larger than the
maximum velocity of the filament.
To better study the movement of the particles, we divide  the particle system
in two groups: the group of particles that are inside the filament by the end
of the simulation, and the group of ejected stars.  
Figure~\ref{fig:trans_vmean} shows the mean velocity of the particles in the
filament as the blue line, the ejected particles as the orange dashed line.
Particles inside the filament clearly follow the gas potential, and move with
the same velocity of the gas.
In contrast, to the ``destroyed'' (see below) case, where the few particles
that manage to stay inside the cluster are the ones  that begin the simulation
moving with velocities close to the one of the filament, in the ``transition''
clusters, the particles inside the filament begin the simulation with lower
mean velocities, but still moving in the same direction of the filament.
When these particles cross the filament, they reach a maximum velocity that is
larger than the maximum velocity of the filament, but remain bounded to the
central part of the potential nonetheless.
On the other hand, particles that are ejected are, on average, moving against
the filament at the moment when the simulation starts.
By the time these particles cross the filament, not only they are moving even
faster than the maximum velocity of the bounded particles when they crossed the
filament, they also reach the filament at a later time, so the difference in
velocity between the stars and the gas is larger than for the previously
mentioned group.
The ejected particles do not cross the filament at the same time, nor with the
same velocity.
Therefore, they will reach different maximum distances from the filament, and
will fall back at different times.
An effect of the different fall back times is that while the particles that reach a
larger distance from the centre of the simulation are starting to fall back
into the filament, particles ejected with a low relative velocity are already
being stirred by the filament and moving in the same direction (although not
with the same velocity or at the same position, so it is unlikely that they
will be recaptured) than the filament, so the mean velocity of this component
of the system will maintain a low value.

\subsection{Destroyed Clusters}\label{sec:dest}

\begin{figure}
    \centering
    \includegraphics{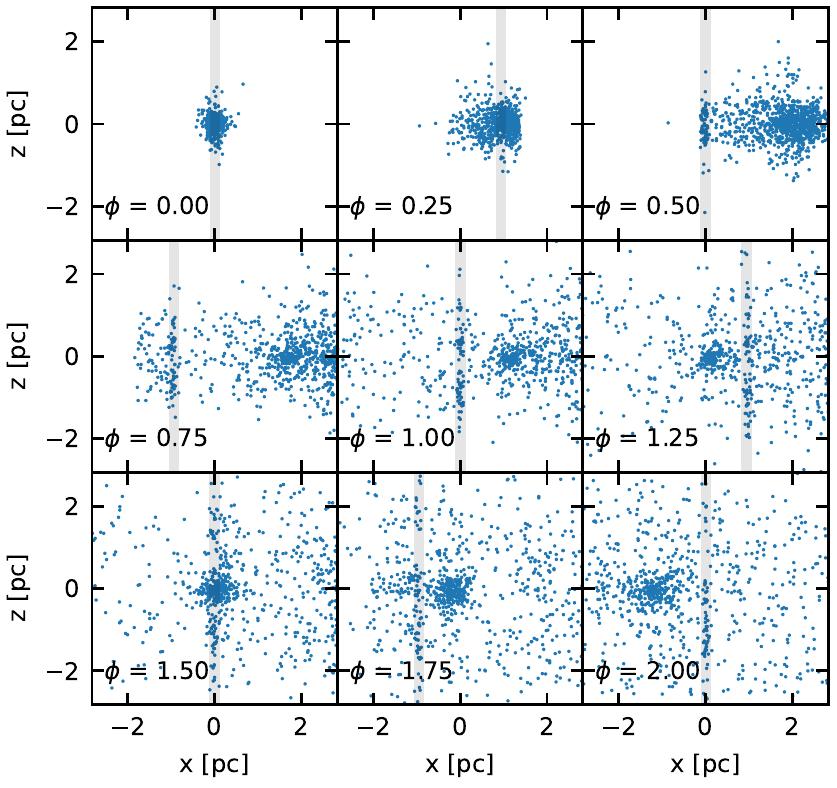}
    \caption{Time evolution of the particles (blue dots), representing stars,
    in an oscillating gas filament (grey vertical line).
    The cluster shown here corresponds to model B ($\Rpl = 0.1$~pc, $\Mpl$
    =500\Msun), inside a filament that oscillates with period P=1.48~Myr and
    amplitude A = 0.94~pc.  
    This produces a remnant belonging to a destroyed cluster (region 3 in
    Figure~\ref{fig:zonesR01})
    The small overdensity, visible from $\phi$~=~1.25 onwards, is in fact two
    smaller clumps, created when the filament went through the main bulk of
    ejected stars at $\phi$~=~1.0. 
    These two associations have opposite $v_y$ velocities, and they do not come
    together to form a new cluster.
    }\label{fig:dest_snap}
\end{figure}

Models in region 3 lose more than 80\% of the particles within the first
oscillation, and half of the initial mass is gone by the time the filament
reaches its maximum distance from the initial position.
The system has lost any resemblance of the original cluster, with streams of
stars trying to catch up with the gas potential and only a few particles still
in the central part of the filament (Figure~\ref{fig:dest_snap}).
After most of the particles are dispersed, there is still a small overdensity
left (Figure~\ref{fig:dest_snap}, middle row), which dissolves after
subsequent filament crossings.

\begin{figure}
    \centering
    \includegraphics{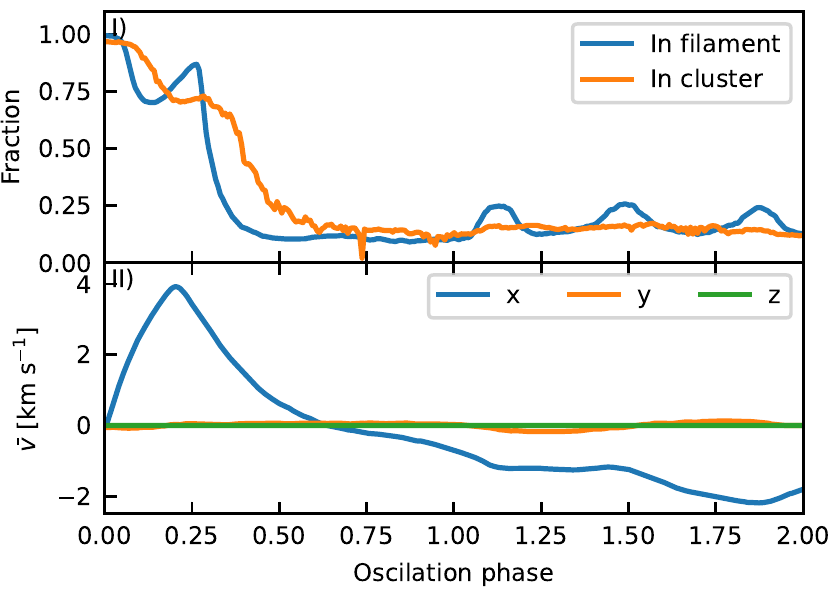}
    \caption{ Same as Figure~\ref{fig:sano_plots}, for Model B and the parameters
    shown in Figure~\ref{fig:dest_snap} (``destroyed'' cluster).
    The cluster shown here corresponds to model B ($\Rpl$=0.1~pc,
    $\Mpl$=500\Msun) , inside a filament that oscillates with period P=1.48~Myr
    and amplitude A = 0.94~pc.
    See caption of Figure~\ref{fig:trans_plots} for a overview of each panel.
    The three bumps in the fraction of particles in the filament
    corresponds to the moment when the filament crosses the bulk of ejected
    particles.
    }\label{fig:dest_plots}
\end{figure}

Figure~\ref{fig:dest_plots}, top panel, shows the catastrophic mass loss at the
beginning of the simulation.
As soon as the filament starts to move, approximately 25\% of the particles
leave the filament, only because they are moving with velocities in the opposite
direction of the movement of the filament.
The next phase of mass loss, bringing the fraction of particles inside the
cluster from $\sim75\%$ to its final  $\sim20\%$, corresponds to the moment
when the filament reaches its maximum distance from its initial position.
As the filament stops moving in the positive direction, and begins to move back
into the centre of the simulation, the cluster is still moving with positive
velocity (Figure~\ref{fig:dest_plots}, lower panel), too fast to be recaptured
by the gas potential, and it will fly past the filament.
Without the potential well of the gas, and stretched by the initial pull of the
filament, the cluster is not able to keep its particles bounded any longer,
leading to a steady decrease of the fraction of particles in the cluster.
The ejected stars do not escape the filament and fall back into it
(Figure~\ref{fig:dest_snap}, middle and bottom rows), explaining the almost
constant fraction of particles in the filament at the end of the simulation in
Figure~\ref{fig:dest_plots}, top panel.

\begin{figure}
    \centering
    \includegraphics{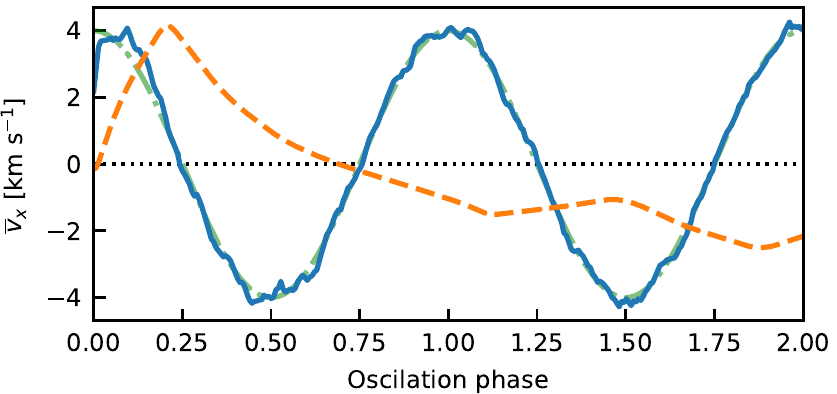}
    \caption{ Mean velocity of the particles in the filament (blue line)
    and of the ejected particles (orange dashed) for the destroyed cluster of
    Figure~\ref{fig:dest_snap}.
    Similar to the behaviour noted in Section~\ref{sec:trans}, the particles in
    the filament have the same velocity than the filament (green dash dotted).
    }\label{fig:dest_vels}
\end{figure}

Since most of the stars have been ejected, the mean velocity of the particles
reflects the motion of the streams around the filament.
Still, some stars move with the filament, so we can separate these two groups
before measuring their mean velocities.
In Figure~\ref{fig:dest_vels}, we select particles that are inside the filament
at the end of the first oscillation and measure the mean velocity of the
selected particles.
It shows that the particles that manage to move with the filament are the ones
that have initial velocities in the $x$ direction close to the velocity of the
filament, in contrast with the ``Filament Associated'' cluster
(Section~\ref{sec:heal}), where nearly all the particles manage to move with
the filament, independently of their initial velocity.
On the other hand, the bulk of the ejected particles re-enter the filament when
the filament has stopped moving in the positive direction and is starting to
turn back towards the centre of the simulation.
At this moment, the particles are moving nearly as fast as the maximum velocity
of the gas potential, and are ejected from the filament.
The peak in velocity of Figure~\ref{fig:dest_vels}, top panel, at
$\phi\sim0.2$, and the secondary knees at $\phi\sim1.1$ and $\phi\sim1.8$ for
the orange dashed line, mark the moment when the ejected particles cross the
filament, which corresponds to the increments of the fraction of particles in
the filament shown in the top panel of Figure~\ref{fig:dest_plots}.

\subsection{Ejected Clusters}\label{sec:ejected}

\begin{figure}
    \centering
    \includegraphics{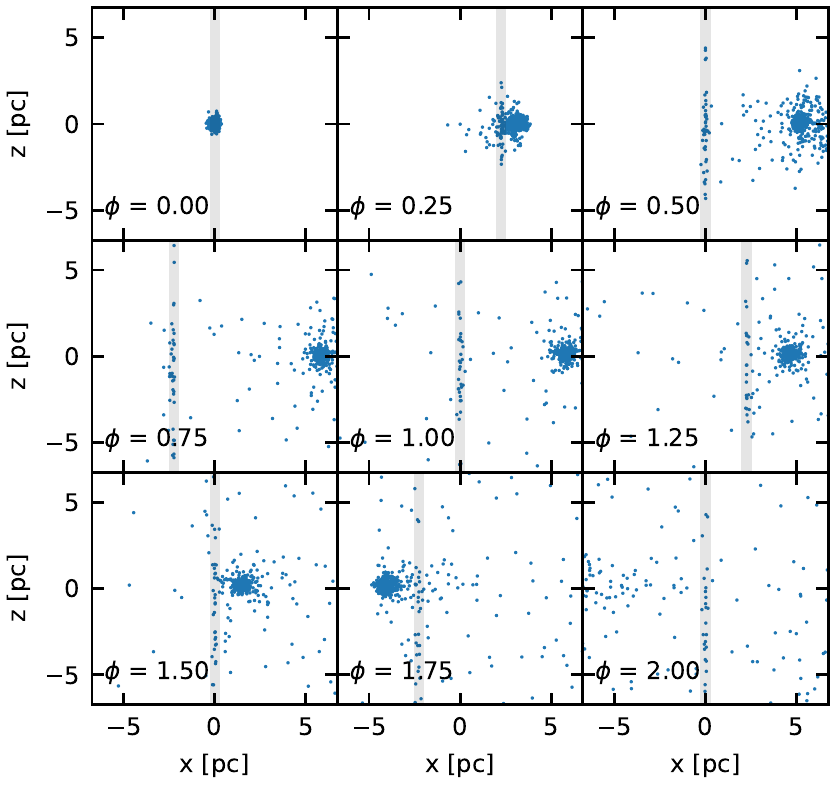}
    \caption{Time evolution of the stars (blue dots) in an oscillating gas
    filament (grey vertical line).
    The cluster shown here corresponds to model C ($\Rpl$=0.1~pc,
    $\Mpl$=1000\Msun), inside a filament that oscillates with period P=3.0~Myr
    and amplitude A = 2.24~pc.
    For this combination of oscillation period and amplitude, the cluster is
    ejected from the filament, reaching distances from the filament large
    enough to leave the plotting area ($\phi$~=~2.0).
    }\label{fig:eye_snap}
\end{figure}

As the name suggests, some clusters are ejected as a whole from the filament. 
That is, under some combinations of filament oscillation amplitude and period,
the cluster escapes filament capture without losing enough mass to disrupt
their equilibrium state.  
In other words, they have a potential well deep enough  to survive removal from
the gas filament. 
These models correspond to region 4 in Figure~\ref{fig:zonesR01}.  
One such example of an ejected cluster (model C, $\Rpl$=0.1~pc,
$\Mpl$=1000~\Msun) is shown in Fig~\ref{fig:eye_snap}.

As mentioned previously, formally our filament has an infinite mass when the
density is integrated to $r \sim \infty$. 
This implies that these ``escaped'' clusters never actually escape the filament.
Inevitably, they will succumb to gravity and fall back into the filament.  
However, in nature, we expect that the filament potential will not behave in
this fashion, and depending on the mass of the filament versus the larger scale
but still local ISM fluctuations these clusters will have a larger survival
probability. 
That said, formally in our results and because of this infinite filament
mass, even the ``escaped'' clusters will never truly escape the grasp of the
filament potential.  
Hence they are inevitably trapped in an auto-destructive cycle of filament
encounters, in which the number of times the cluster suffers a close and
destructive filament encounter depends on the velocity of the cluster relative
to the filament at the moment of first separation from the filament.  
Regardless of what happens in nature versus the artificial simulations, if the
relative velocity is small, the cluster will on short order fall back into the
filament, suffering multiple encounters and accompanying severe disruption of
its structure. 
On the other hand, a high relative velocity at first separation will keep the
cluster safe from filament-induced destruction for a longer time because of a
reduced number of damaging encounters with the filament potential per unit
time, or full escape from the local filament potential.

\begin{figure}
    \centering
    \includegraphics{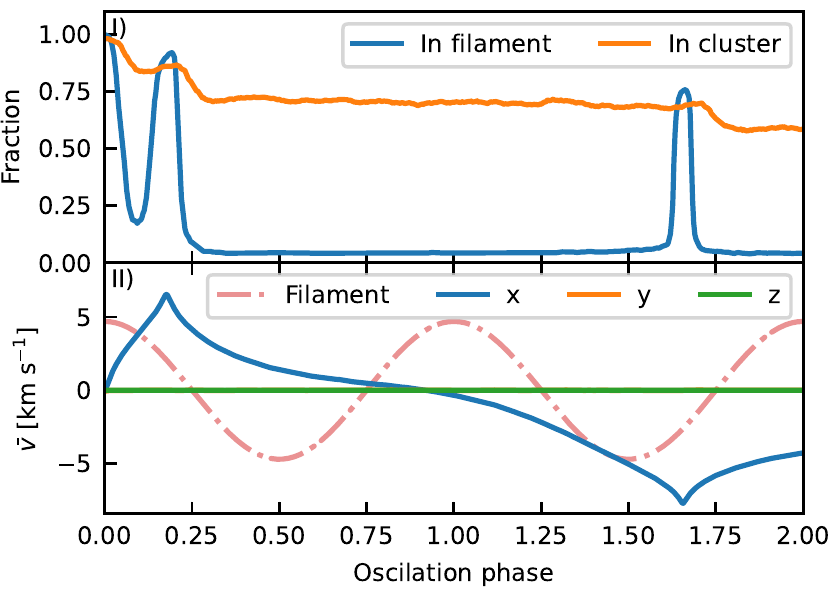}
    \caption{ Same as Figure~\ref{fig:sano_plots}, but for Figure~\ref{fig:eye_snap},
    ``ejected'' cluster.
    The cluster shown here corresponds to model C ($\Rpl = 0.1$~pc, $\Mpl =
    1000$\Msun), inside a filament that oscillates with period P = 3.0 Myr and
    amplitude A = 2.24~pc, which ejects the cluster at an oscillation phase of
    $\phi\sim0.2$.
    }\label{fig:eye_plots}
\end{figure}

Similar to the clusters that fall in the category ``destroyed''
(Section~\ref{sec:dest}), Figure~\ref{fig:eye_plots} shows that ``ejected''
clusters also go through an initial mass loss phase when the simulation begins
(top panel) as the cluster starts to gain velocity to catch up with the moving
filament.
Particles in front of the filament, and with negative velocities, i.e.\ at the
right side of the filament and moving to the left side in
Figure~\ref{fig:eye_snap}, are the first particles that leave the cluster and
are left behind by the filament.
This represents around 10\% of the initial mass of the cluster, and it is lost
to the cluster before the filament reaches its maximum distance from the origin.
There is a second phase of mass loss near the beginning of the simulation that
corresponds to the moment when the cluster leaves the filament for the first
time.
This time, the particles that leave the cluster are the particles that begin
the simulation with velocities close to the velocity of the filament; these
particles stay inside the filament and are stripped from the cluster once the
filament starts to  recede back to its initial position.
They follow the inevitable pull of the filament and are bound to it.
After the second ejection, the cluster is outside the filament and moving away
from it.
As long as the cluster is outside the filament, it will not lose any more mass.
Eventually, the cluster falls back into the filament, and will lose a fraction
of its particles each time this happens.

As with the previous cases, the movement of the filament along the $x$ axis
induces the star system to have a mean velocity along the $x$ direction.
The cluster shown in Figure~\ref{fig:eye_snap} accelerates from rest until it
crosses the filament, shortly before the filament reaches its maximum distance
from the centre of the simulation.
At this moment, the cluster is moving at $\sim7.0$\kms
(Figure~\ref{fig:eye_plots}, bottom panel), $\sim 4.5$\kms\ faster than the
filament. 
The crossing happens near the turn and, in consequence, the gas potential
will not be able to slow down the cluster enough to recapture it, so the star
system will overtake the filament.
The cluster is able to reach a velocity larger than the maximum velocity of the
filament.
This can be explained with an observer moving with the filament.
In the simulation frame, the cluster starts at rest, with
$v_\mathrm{cl}=0$~\kms, while the filament is moving with $v_\mathrm{fil}(t=0)
= v_\mathrm{max}$.
For the observer in the filament, at $t=0$, it is the cluster what its moving
away, with $v_\mathrm{cl}=-v_\mathrm{max}$, while the filament is static.
As the simulation continues, eventually the cluster stops moving away and
starts to fall back into the centre of the filament.
At the moment of the filament crossing, the cluster will be moving with
$v_\mathrm{cl}=v_\mathrm{max}+\delta v$, where $\delta v$ is caused by the
fictitious forces due to the acceleration of the filament in the simulation
frame.
Back in the simulation frame, the velocity of the cluster will be
$v_\mathrm{max}+v_\mathrm{fil}(t_\mathrm{encounter}) + \delta v$, which will be
larger than $v_\mathrm{max}$, unless the encounter happens at a moment when
$\delta v < -v_\mathrm{fil}(t_\mathrm{encounter})$.
Then, as the filament recedes, the cluster starts to slow down, and reaching a
maximum distance of 5.9~pc from its initial position.
Once the cluster stops, it starts to fall back into the filament.
Eventually, there is a new interaction between the gas potential and the star
system, moment at which the cluster is, again, moving too fast to be captured
by the filament. In this second interaction, the difference in velocity between
the cluster and the filament is larger than in the first interaction: with the
cluster moving at $\sim8.7$\kms, the velocity relative to the filament is
$\sim6.1$\kms.
This suggests that an ejected cluster could not only not be recaptured by the
filament, it might gain enough velocity to escape the gas cloud that created
it after a few crossings.

\subsection{Orion Nebula Cluster}

\begin{figure}
    \includegraphics{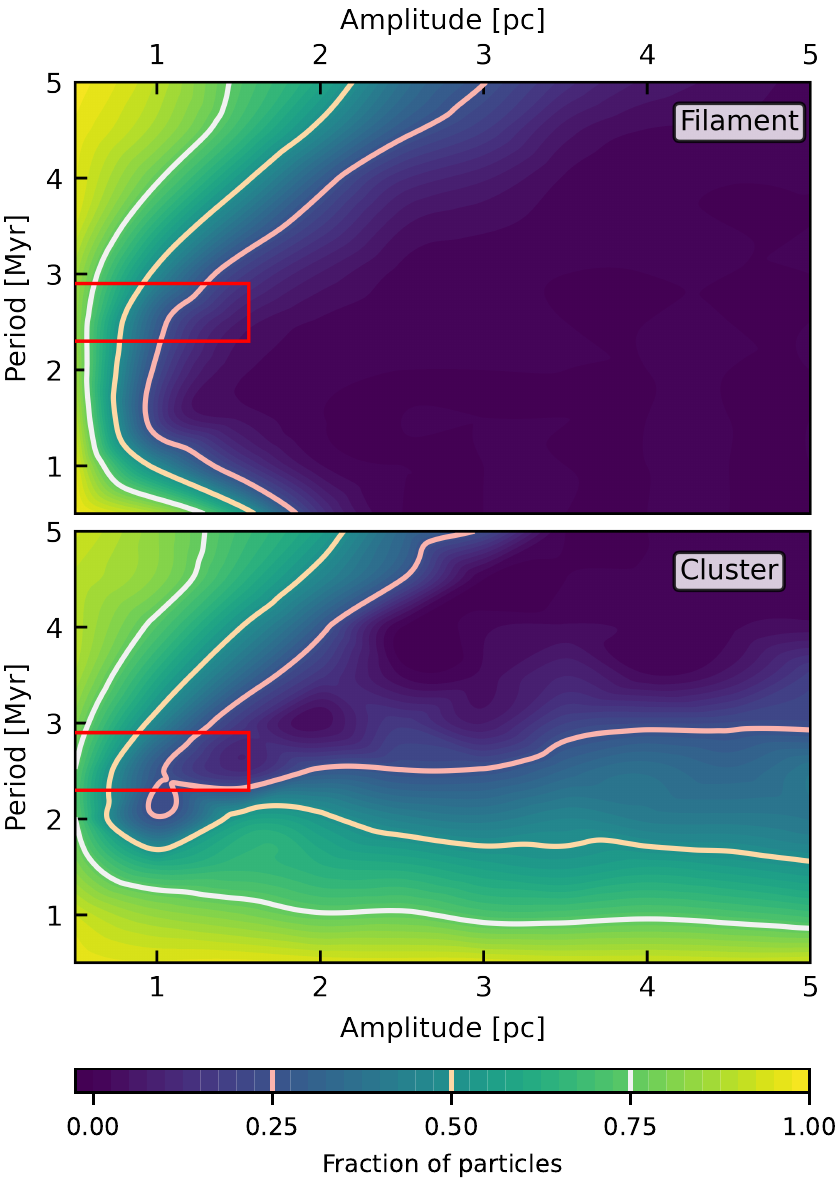}
    \caption{Fractions of particles in the filament (top) and in the cluster
    (bottom) for a model with radius and mass similar to the ONC.
    The red rectangle shows the possible oscillation parameters of the ISF,
    while the contours represent fractions of 25\%, 50\% and 75\% of the initial
    particles.
    }\label{fig:onc}
\end{figure}

As our simulations have shown, the end state of the cluster depends on both, the
parameters of the cluster and the parameters of the oscillation of the
filament.
For the case of the ONC, \citet{stutz_2018} show that the
cluster can be modelled as a Plummer sphere with a total mass of 1124~\Msun\,
and a Plummer radius of 0.36~pc.
The fraction of particles left in the filament and in the cluster after one
filament oscillation is shown in Figure~\ref{fig:onc}, where we see that,
indeed, there are regions in the period-amplitude plot that result in a
destroyed cluster.
From the different pairs of period and amplitude, of special interest are the
values that are close to the estimated oscillation parameters of the ISF
\citep{stutz_2016, boekholt_2017, dominik_2018, stutz_2018a}, shown in
Figure~\ref{fig:onc} in the red rectangle.
This suggests that, unless the oscillation has an amplitude smaller than the
radial extent of the cluster, the ONC will lose a considerable fraction of
stars and will be destroyed by the filament.

Observations of the ONC indicate that the cluster is in expansion
\citep{jones_1988, kroupa_1999, scally_2005} and located slightly in the
foreground of the ISF \citep{hillenbrand_1998,odell_2001}.
Under the slingshot scenario, these observations could indicate that the
filament is close to its turnaround, and the cluster is being ejected from the
filament, as already pointed out by \citet{stutz_2018}.
Removing the gas from the cluster lowers the potential that is keeping the
stars bounded together, causing the expansion of the cluster. 
Once outside the filament, the cluster is dissolved by the time the filament
completes its second oscillation, so we expect a bleak future for the ONC, with
this cluster being transformed to a smaller association of stars, or completely
dispersing into the Galactic field.

\section{Discussion and Conclusions}\label{sec:conc}

In this work we investigate the effects of an oscillating gas filament, also
known as the Slingshot model, in the early evolution of a young star cluster.
We achieve this via numerical simulations of a spherical distribution of
particles inside a cylindrically symmetrical gas potential that oscillates with
a sinusoidal motion.
We are able to simulate the kinematics of the system by coupling an N-body
solver with an analytical background potential.

Clusters in oscillating filaments will lose particles as soon as the simulation
starts due to tides produced by the motion of the filament.
The amount of ejected particles depends on the parameters of the cluster, the
amplitude of the oscillation and the period of the oscillation.

The motion of the filament will cause the ejection of stars from the cluster.
The majority of the ejected stars leave the filament when the filament reaches
its maximum amplitude for the first time.
The ejected particles move in the direction of the motion of the filament and
eventually fall back into the filament.

We identify four outcomes for the cluster under the motion of the filament.
We dub them as ``Filament Associated'', ``transition'', ``destroyed'' and
``ejected'' clusters, depending on the fraction of particles left inside the
cluster and inside the filament.
``Filament Associated Cluster'' correspond to the clusters that remain inside
the filament and keep at least 80\% of their particles.
On the other extreme, ``destroyed'' clusters  are the remnants that keep at
most 20\% of their particles inside the filament and inside $5\Rpl$ from the
centre of density of the star system.
In the middle we have ``transition'' clusters, where a significant fraction of
particles leave the cluster but there is still a clear overdensity of stars in
the filament.
Finally, ``ejected'' are, as the name implies, the clusters that are ejected by
the motion of the filament and are not destroyed by the removal of the
background potential.
The fate of the cluster in an oscillating filament is decided quickly.
By the time the filament reaches its maximum distance from its initial position,
any star that manages to stay inside the cluster or the gas filament will
likely stay there for the rest of the simulation time.

In a real cluster, stars would have different masses, which gives rise to
processes of relaxation and mass segregation.
As indicated in Section~\ref{subsec:cl}, we use equal mass particles in our
simulations, so we do not observe such process.
Although using a mass spectrum could improve our results, the effects of the
filament on the cluster take place in timescales shorter than the relaxation
time of the clusters (Table~\ref{tab:init}).
In the filaments with the largest period, the ejection or destruction of the
cluster happens during the first 1.25~Myr (see Section~\ref{subsec:4zone}),
while the cluster with the shortest relaxation time, model C, has
$t_\mathrm{cr}=1.21$~Myr.
For filaments with shorter periods, the cluster does not have enough time to
relax before the motion of the filament stirs the stars.
On the other hand, \citet{mouri_2002} shows that the timescales for mass
segregation are  shorter than the relaxation time of the system, so mass
segregation could be important in young systems. 
Observations show \citep{hillenbrand_1998, allison_2009, pang_2013,
plunkett_2018, dib_2019} that star clusters either form in a mass segregated
state or reach mass segregation in a short timespan. 
In particular, the ONC is already mass segregated \citep{hillenbrand_1998}.
Since we consider that relaxation effects are not important in our simulations,
and the study of mass segregation under the Slingshot model is outside the
scope of this paper, we did not add a mass spectrum for our particles.
Nonetheless, future numerical simulations could explore the effects of the
Slingshot on the mass segregation process of young clusters.

According the \citet{stutz_2018}, the distribution of stars in the ONC 
is well characterized by a Plummer profile with a Plummer mass of 1124~\Msun\,
and a Plummer radius of 0.36~pc.
A cluster with said parameters can be destroyed if the oscillation amplitude of
the filament is larger than the radial extent of the cluster, given an
estimated oscillation period of $\sim2.5$ Myr.

\section*{Acknowledgements}

The authors like to thank the anonymous referee for their comments, which
helped to significantly improve the paper.
DRMC, MCBMI and MF acknowledge financial support from Fondecyt regular No.
1180291. 
DRMC acknowledges funding through a beca Conicyt doctorado nacional convocatoria 2020.
MF also acknowledges support by Conicyt Quimal No. 170001, and the ANID
BASAL projects ACE210002 and FB210003. 
TCNB was supported by funds from the European Research Council (ERC) under the
European Union’s Horizon 2020 research and innovation program under grant
agreement No 638435 (GalNUC).
AS gratefully acknowledges support from the ANID BASAL projects ACE210002 and FB210003.  
AS acknowledges support from the Fondecyt Regular (project code 1220610).

\section*{Data Availability}
The data underlying this article will be shared on reasonable request to the
corresponding author.


\bibliographystyle{mnras}
\bibliography{docs} 







\bsp	
\label{lastpage}
\end{document}